\documentclass[11pt]{article}

\usepackage[a4paper,margin=1in]{geometry}
\usepackage{amsmath,amssymb,amsfonts,mathtools,bm}
\usepackage{graphicx}
\usepackage{hyperref}
\usepackage{enumitem}
\usepackage{booktabs}
\usepackage{microtype}
\usepackage{authblk}
\usepackage{orcidlink}

\hypersetup{
  colorlinks=true,
  linkcolor=blue,
  citecolor=blue,
  urlcolor=blue
}

\newcommand{\mpl}{m_{\mathrm{Pl}}}
\newcommand{\Epl}{E_{\mathrm{Pl}}}

\newcommand{\dd}{\mathrm{d}}

\title{Black-hole thermodynamics in doubly special relativity: near-horizon $g/f$ temperature scaling under a shared operational scale}

\author[1]{Abdelmalek Boumali\,\orcidlink{0000-0003-2552-0427}\thanks{Email: \texttt{boumali.abdelmalek@gmail.com}}}
\author[2]{Nosratollah Jafari\,\orcidlink{0000-0002-0285-6107}\thanks{Email: \texttt{nosrat.jafari@fai.kz}}}
\affil[1]{Echahid Cheikh Larbi Tebessi University, Tebessa, Algeria}
\affil[2]{Fesenkov Astrophysical Institute, Almaty, Kazakhstan}

\date{\today}

\begin{document}
\maketitle

%============================================================
% Abstract
%============================================================
\begin{abstract}
Doubly Special Relativity (DSR) deforms special-relativistic kinematics while preserving a relativity
principle by introducing, in addition to the invariant speed of light, a second invariant scale that is
typically identified with the Planck energy $E_{\mathrm{Pl}}$. In curved spacetime, however, the physical
meaning of the ``energy'' entering DSR-inspired modified dispersion relations (MDRs) is ambiguous:
Killing energies, local observer energies, and phase-space invariants need not coincide. Consequently,
for black-hole applications, this ambiguity can be just as important as the algebraic form of the MDR
itself.

The main objective of this paper is to clarify, within a single and consistent framework, whether two
widely used curved-spacetime implementations of DSR deformations really lead to different Hawking-temperature
predictions or merely represent different parametrizations of the same near-horizon effect. To this end,
we compare (i) MDRs imposed in local orthonormal frames on an energy-independent background geometry and
(ii) the rainbow-metric proposal, in which the deformation is encoded in an energy-dependent family of
effective metrics. Restricting to static, spherically symmetric horizons and, moreover, evaluating both
constructions at the same finite operational scale $E_\star$, we show that the two prescriptions lead to
the same near-horizon temperature rescaling,
\begin{equation}
T(E_\star)=T_0\,\frac{g(E_\star/\Epl)}{f(E_\star/\Epl)}\,,\qquad T_0=\frac{\kappa_0}{2\pi}\,.
\end{equation}
Therefore, our result is not a first-principles derivation of a universal DSR prediction; rather, it is a
controlled equivalence statement showing that, under a shared operational prescription, the local-MDR and
rainbow descriptions amount to the same parametrization of the near-horizon temperature shift.

We then illustrate this statement for an Amelino--Camelia-type MDR and for the original Magueijo--Smolin
DSR invariant, for which $f=g$ and therefore $T(E_\star)=T_0$ at this level. Next, we extend the
comparison to a two-parameter generalized DSR family (G-DSR/GDRS) characterized at leading order by
$(\alpha_2,\Delta\alpha)$, obtaining
\begin{equation}
T_{\mathrm{GDRS}}(E_\star)
=
T_0\sqrt{\frac{1-2\,\Delta\alpha\,(E_\star/\Epl)}{1-2\,\alpha_2\,(E_\star/\Epl)}}
\simeq
T_0\!\left[1-(\Delta\alpha-\alpha_2)\frac{E_\star}{\Epl}\right].
\end{equation}
In this way, we make explicit that the leading correction is controlled by the single combination
$\Delta\alpha-\alpha_2$, which vanishes on the symmetric subfamily $\Delta\alpha=\alpha_2$. Finally, we
show how the size of the correction depends on the prescription used for $E_\star$: for the common choice
$E_\star\sim T_0$, the relative shift is suppressed by $1/(M\Epl)$ and is therefore negligible for
macroscopic black holes. Accordingly, observable or phenomenological consequences must come either from
Planckian regimes or from ingredients beyond the near-horizon temperature, such as greybody factors,
phase-space measures, thresholds, and deformed multi-particle composition laws.
\end{abstract}

\vspace{0.5em}

%============================================================
\section{Introduction}
%============================================================
Hawking's discovery that black holes radiate \cite{Hawking1975}, together with the Bekenstein area law for
entropy \cite{Bekenstein1973}, established a thermodynamic interpretation of horizon mechanics in which the
Hawking temperature is set by the surface gravity, $T_0=\kappa_0/(2\pi)$ (in units $G=c=\hbar=k_B=1$).
Comprehensive treatments may be found in \cite{WaldBook,PoissonToolkit}. Black-hole thermodynamics thus
provides a constrained setting for testing how semiclassical quantum field theory responds to ultraviolet
(UV) modifications.

A standard motivation for Planck-scale deformations is the trans-Planckian problem: in conventional
Hawking calculations, outgoing quanta originate from modes that acquire arbitrarily large local
frequencies when traced backwards toward the horizon. A large body of work nevertheless suggests that the
late-time Hawking flux is rather robust under many UV deformations of the microphysics, including broad
classes of modified dispersion relations and analogue-gravity constructions
\cite{Jacobson1991,Unruh1995,Brout1995,CorleyJacobson1996,BarceloLiberatiVisser2011}. Even so, MDR-based
effective descriptions remain useful because they can modify the inferred near-horizon temperature, the
propagation of wave packets, and the detailed spectrum through group velocities, densities of states, and
greybody factors.

Doubly Special Relativity (DSR) is an especially interesting framework in this context because it aims to
deform relativistic kinematics while preserving a relativity principle through the introduction of a second
invariant scale, typically the Planck energy $\Epl$
\cite{AmelinoCamelia2001,AmelinoCamelia2002,MagueijoSmolin2002,KowalskiGlikman2005,AmelinoCamelia2010}.
In flat spacetime, MDRs are then accompanied by non-linear Lorentz transformations and, in general,
deformed composition laws for multi-particle systems. In curved spacetime, however, one immediately faces a
serious operational problem: the energy entering the deformation is not uniquely defined. One may use a
conserved Killing energy at infinity, a local energy measured by static or freely falling observers, or a
phase-space invariant. Different choices need not agree numerically, and if they are mixed
inconsistently they can effectively reintroduce a preferred frame.

This ambiguity is central to the comparison between the two curved-spacetime prescriptions that dominate
much of the DSR/rainbow black-hole literature. In the first, the spacetime geometry is kept classical and
energy independent, while particle propagation is governed by an MDR imposed in a local orthonormal frame.
In the second, the deformation is encoded in an energy-dependent family of effective metrics, as in rainbow
gravity \cite{MagueijoSmolinRainbow2004}. Against this background, the main objective of the present work
is to determine whether these two procedures genuinely yield different Hawking-temperature predictions or,
instead, become equivalent once the same physical energy prescription is adopted. More precisely, we show
that for static, spherically symmetric horizons, and provided that the same finite operational scale
$E_\star$ is used in both constructions, they lead to the same near-horizon temperature rescaling by the
ratio $g/f$.

We emphasize from the outset that this is a conditional equivalence statement, not a derivation of a
unique DSR prediction from first principles. Rather, what is being shown is that once a common prescription
for the deformation scale is imposed, the apparent difference between ``local MDR'' and ``rainbow metric''
descriptions largely reduces to parametrization at the level of the tunneling/surface-gravity temperature.
Therefore, this framing is important because the operational definition of $E_\star$ remains the dominant
source of ambiguity in curved-spacetime DSR applications.

Besides the standard Amelino--Camelia (AC) and Magueijo--Smolin (MS) realizations, we also incorporate a
two-parameter generalized DSR family (G-DSR/GDRS) with leading-order parameters
$(\alpha_2,\Delta\alpha)$ \cite{BoumaliJafariChargui2026}. In this way, the analysis makes transparent
which combination of deformation parameters controls the leading correction and which subfamily leaves the
near-horizon temperature unchanged.

The paper is organized as follows. Section~\ref{sec:dsr_flat} reviews the MDR conventions and the
representative DSR realizations used in the comparison. Section~\ref{sec:dsr_curved} discusses
curved-spacetime extensions and formulates the energy-scale ambiguity precisely. Section~\ref{sec:temperature_scaling}
derives the common $g/f$ rescaling for static horizons using both local-frame MDR and rainbow-metric
viewpoints. Section~\ref{sec:examples} applies the result to the AC, MS, and G-DSR/GDRS cases and examines
simple Schwarzschild implications, including the parametric size of the correction. Section~\ref{sec:beyond_temperature}
explains why agreement in the near-horizon temperature does not imply identical evaporation phenomenology,
and Section~\ref{sec:gdrs_discussion} discusses the interpretation and limitations of the generalized DSR
extension. We conclude with a summary of what has and has not been established.

%============================================================
\section{DSR in flat spacetime: MDR conventions and representative models}
\label{sec:dsr_flat}
%============================================================

\subsection{Clarifying the MDR input: functional form versus operational energy}
An MDR provides an effective, Planck-suppressed deformation of the special-relativistic mass shell. In DSR
this deformation is usually accompanied by a non-linear realization of Lorentz transformations in momentum
space, designed to preserve a relativity principle in the presence of an invariant high-energy scale
\cite{AmelinoCamelia2002,KowalskiGlikman2005}. In many curved-spacetime applications, however, the MDR is
used as a phenomenological one-particle input, and the principal ambiguity shifts from the algebraic form
of the MDR to the operational meaning of the energy that enters it.

In this work we adopt the standard $f$--$g$ parametrization,
\begin{equation}
E^2\,f^2\!\left(\frac{E}{\Epl}\right)-p^2\,g^2\!\left(\frac{E}{\Epl}\right)=m^2,
\label{eq:fg_mdr}
\end{equation}
with $f(0)=g(0)=1$ ensuring the low-energy recovery of special relativity. For massless particles,
\eqref{eq:fg_mdr} reduces to
\begin{equation}
E\,f\!\left(\frac{E}{\Epl}\right)=p\,g\!\left(\frac{E}{\Epl}\right)\qquad (m=0),
\label{eq:massless_relation}
\end{equation}
so that the ratio $g/f$ controls the deformation of the on-shell relation between energy and momentum.
Depending on the explicit model, the deformation may also imply an energy-dependent group velocity and, in
DSR, typically a modified composition law for multi-particle states \cite{Hossenfelder2007}.

A crucial point for black-hole applications is that the argument $E/\Epl$ in \eqref{eq:fg_mdr} must be
interpreted operationally. In flat spacetime this is unambiguous (up to standard inertial-frame
conventions), but in curved spacetime one may consider the conserved energy associated with a Killing
vector, energies measured by a specified class of observers, or invariants associated with phase-space
geometry \cite{GirelliLiberatiSindoni2007,BarcaroliPfeifer2015}. The comparison between local-frame MDR and
rainbow-metric computations performed below is meaningful only once a common choice of operational scale is
specified.

\subsection{Amelino--Camelia (AC) type MDR}
A commonly used AC-type MDR, truncated at leading order in $\Epl^{-1}$, reads
\cite{AmelinoCamelia2001,AmelinoCamelia2002}
\begin{equation}
E^2-p^2-m^2+\eta\,\frac{E}{\Epl}\,p^2=0\,,
\label{eq:ac_mdr}
\end{equation}
where $\eta$ is dimensionless. Rearranging yields
\begin{equation}
E^2-p^2\Bigl(1-\eta \frac{E}{\Epl}\Bigr)=m^2.
\label{eq:ac_rearranged}
\end{equation}
Matching \eqref{eq:ac_rearranged} to \eqref{eq:fg_mdr} provides a convenient identification (exact in $x$)
\begin{equation}
f_{\rm AC}(x)=1,\qquad g_{\rm AC}(x)=\sqrt{1-\eta x}\,,
\qquad x\equiv \frac{E}{\Epl}.
\label{eq:fg_AC_exact}
\end{equation}
For $\eta>0$ the effective spatial factor $g$ decreases with energy, and the MDR becomes ill-defined for
$x\ge 1/\eta$ in this truncated parametrization; such features should be interpreted within the validity
domain of the effective description.

\subsection{Magueijo--Smolin (MS) DSR invariant}
In the Magueijo--Smolin realization, an invariant energy scale is implemented through a non-linear
representation of Lorentz transformations \cite{MagueijoSmolin2002}. In commonly used variables, the
modified invariant reads
\begin{equation}
\frac{\eta^{ab}p_a p_b}{(1-E/\Epl)^2}=m^2,
\label{eq:ms_invariant}
\end{equation}
equivalently,
\begin{equation}
\frac{E^2-p^2}{(1-E/\Epl)^2}=m^2
\quad\Longleftrightarrow\quad
E^2-p^2=m^2\,(1-E/\Epl)^2.
\label{eq:ms_mdr}
\end{equation}
In the parametrization \eqref{eq:fg_mdr} this corresponds to
\begin{equation}
f_{\rm MS}(x)=g_{\rm MS}(x)=\frac{1}{1-x}\,,
\label{eq:fg_MS}
\end{equation}
so that $g_{\rm MS}/f_{\rm MS}=1$. This equality will play an important role below: even though the MDR is
non-trivial, the near-horizon temperature correction vanishes in the common $g/f$ scaling.

\subsection{Generalized DSR (G-DSR/GDRS): a two-parameter leading-order family}
\label{sec:gdrs_flat}
A convenient leading-order generalization that captures broad classes of deformed kinematics is the
two-parameter G-DSR/GDRS framework \cite{BoumaliJafariChargui2026}. Its MDR can be written as
\begin{equation}
E^2-p^2 - 2\alpha_2\,\frac{E^3}{\Epl} + 2\,\Delta\alpha\,\frac{E\,p^2}{\Epl} = m^2,
\label{eq:gdrs_mdr_raw}
\end{equation}
where $\alpha_2$ and $\Delta\alpha\equiv \alpha_3-\alpha_1$ are dimensionless. Factoring the leading
corrections gives the suggestive form
\begin{equation}
E^2\bigl(1-2\alpha_2 x\bigr)-p^2\bigl(1-2\Delta\alpha x\bigr)=m^2,
\qquad x\equiv \frac{E}{\Epl},
\label{eq:gdrs_mdr_factored}
\end{equation}
so that matching to \eqref{eq:fg_mdr} yields
\begin{equation}
f_{\rm GDRS}(x)=\sqrt{1-2\alpha_2 x},
\qquad
g_{\rm GDRS}(x)=\sqrt{1-2\Delta\alpha x}.
\label{eq:fg_GDRS}
\end{equation}

At leading order, $\alpha_2$ controls the deformation associated with the energy sector ($E^2$ term), while
$\Delta\alpha$ controls the deformation associated with the spatial momentum sector ($p^2$ term). In static
black-hole applications where the Hawking temperature depends only on $g/f$, the difference
$\Delta\alpha-\alpha_2$ becomes the relevant combination controlling the leading correction. Importantly,
the symmetric subfamily $\Delta\alpha=\alpha_2$ has $f=g$ and therefore shares with the MS realization the
property that the tunneling/surface-gravity temperature remains unmodified at this level.

At leading order, the AC form \eqref{eq:ac_rearranged} is recovered by setting $\alpha_2=0$ and
$2\Delta\alpha=\eta$, so that $g_{\rm GDRS}(x)=\sqrt{1-\eta x}$. The MS invariant \eqref{eq:fg_MS} is an
all-orders symmetric choice with $f=g$; similarly, the GDRS family contains a symmetric branch
$\Delta\alpha=\alpha_2$ for which $g/f=1$ at leading order, illustrating that a non-trivial MDR does not
necessarily imply a modified Hawking temperature.

For convenience, Table~\ref{tab:fg_models} collects the $f$--$g$ functions and the corresponding
temperature ratio $g/f$ for the models considered.
\begin{table}[t]
\centering
\begin{tabular}{@{}llll@{}}
\toprule
Model & $f(x)$ & $g(x)$ & $g(x)/f(x)$ \\
\midrule
AC & $1$ & $\sqrt{1-\eta x}$ & $\sqrt{1-\eta x}$ \\
MS & $(1-x)^{-1}$ & $(1-x)^{-1}$ & $1$ \\
G-DSR/GDRS & $\sqrt{1-2\alpha_2 x}$ & $\sqrt{1-2\Delta\alpha x}$ &
$\sqrt{\dfrac{1-2\Delta\alpha x}{1-2\alpha_2 x}}$ \\
\bottomrule
\end{tabular}
\caption{Representative deformations in the $f$--$g$ parametrization with $x=E/\Epl$.}
\label{tab:fg_models}
\end{table}

%============================================================
\section{DSR in curved spacetime: implementations and the energy-scale issue}
\label{sec:dsr_curved}
%============================================================
DSR is formulated as a deformation of flat-spacetime kinematics; consequently, its extension to curved
spacetime is not unique and depends on which structures are taken to retain observer-independent meaning
in the presence of gravity. For black-hole thermodynamics, the primary challenge is to couple deformed
momentum-space structures to a spacetime geometry without smuggling in a preferred frame through an
inconsistent energy identification. Three approaches are especially relevant: local-frame MDRs on a fixed
background, phase-space/Hamilton geometry and relative locality, and rainbow gravity.

\subsection{Local-frame MDRs on a fixed background}
A pragmatic strategy retains a classical, energy-independent spacetime metric $g_{\mu\nu}(x)$ while imposing
an MDR in local orthonormal frames. Introducing tetrads $e^{a}{}_{\mu}(x)$ such that
$g_{\mu\nu}=\eta_{ab}\,e^{a}{}_{\mu}e^{b}{}_{\nu}$, one defines locally measured components
$p_a=e_a{}^{\mu}p_\mu$ and imposes a constraint $\mathcal{C}(p_a;\Epl)=m^2$ pointwise. In WKB treatments,
$p_\mu=\partial_\mu S$ with $S$ the Hamilton--Jacobi action, and the MDR acts as a Hamiltonian constraint
governing characteristic curves.

This prescription preserves standard geometric notions of horizon, surface gravity, and Killing symmetries
while encoding Planck-scale effects in the matter sector. Its main ambiguity, however, lies in specifying
the \emph{finite} energy scale at which the MDR functions are to be evaluated. In particular, for static
black holes, local energies measured by stationary observers diverge at the horizon. Meaningful MDR
applications must therefore either employ regular frames (e.g.\ freely falling orthonormal frames) or
explicitly postulate that the deformation functions are evaluated at a finite operational scale
$E_\star$ associated with the emitted quanta (as we do in this work).

\subsection{Phase-space geometry and relative locality}
A more structural viewpoint interprets DSR as curvature of momentum space, with an accompanying
observer-dependence of locality (``relative locality'') \cite{RelativeLocality2011}. Hamilton-geometry
approaches view MDRs as defining geometric structures on the cotangent bundle $T^\ast\mathcal{M}$ and study
their implications for propagation and observables \cite{BarcaroliPfeifer2015}. In specific cases,
MDR-induced constructions can be related to Finsler-like geometries \cite{GirelliLiberatiSindoni2007}. From
this perspective, the identification of the deformation scale in curved spacetime is part of the model's
physical content: the ``energy'' entering the MDR should be tied to the phase-space structure rather than
chosen ad hoc. While these approaches are conceptually appealing, they can be technically more involved;
our focus here is on the two prescriptions most commonly used in black-hole thermodynamics calculations.

\subsection{Rainbow gravity}
In rainbow gravity, the deformation is encoded through energy-dependent orthonormal frames
\cite{MagueijoSmolinRainbow2004}
\begin{equation}
e^{0}(E_\star)=\frac{\tilde e^{0}}{f(E_\star/\Epl)},\qquad
e^{i}(E_\star)=\frac{\tilde e^{i}}{g(E_\star/\Epl)}\,,
\label{eq:rainbow_frames}
\end{equation}
which induce a one-parameter family of effective metrics
$g_{\mu\nu}(E_\star)=\eta_{ab}e^{a}{}_\mu(E_\star)e^{b}{}_\nu(E_\star)$. In the black-hole context, this
framework has been used to discuss modifications of the Hawking temperature and, in some model choices,
the possibility of evaporation remnants \cite{LingHuLi2006,AliFaizalKhalil2014,GalanMenaMarugan2006}.

Conceptually, rainbow gravity should be interpreted as an effective description: a given probe of energy
$E_\star$ ``sees'' an effective geometry. This raises well-known questions concerning the implementation of
the equivalence principle and the consistency of multi-particle descriptions, issues that are connected
in DSR to the non-trivial structure of multi-particle kinematics \cite{Hossenfelder2007} and to locality
considerations \cite{Hossenfelder2010}. In the present work, we use rainbow gravity in its most conservative
role: as a parametrization of MDR effects in near-horizon computations, with a fixed operational choice of
$E_\star$.

\subsection{The central ambiguity: which energy enters $E_\star/\Epl$?}
In asymptotically flat black-hole spacetimes, it is essential to distinguish the conserved Killing energy
at infinity, $E_\infty$, from energies measured locally by specific observers. For a generic static,
spherically symmetric geometry,
\begin{equation}
\dd s^2=-F(r)\,\dd t^2+\frac{\dd r^2}{F(r)}+r^2\,\dd\Omega^2,
\label{eq:static_metric}
\end{equation}
a static observer at radius $r$ measures
\begin{equation}
E_{\rm stat}(r)=\frac{E_\infty}{\sqrt{F(r)}}\,,
\label{eq:static_energy}
\end{equation}
which diverges as $r\to r_h$ because $F(r_h)=0$. By contrast, orthonormal frames associated with freely
falling observers can remain regular at the horizon, and the corresponding locally measured energy for a
quantum of finite $E_\infty$ can remain finite. Consequently, any DSR-motivated curved-spacetime extension
must specify whether the deformation depends on $E_\infty$, on a local energy measured in a chosen regular
frame, or on a phase-space invariant.

To facilitate a transparent comparison between prescriptions, we adopt a standard operational assumption
used in much of the tunneling/rainbow literature: the deformation functions $f$ and $g$ are evaluated at a
\emph{finite} physical scale $E_\star$ associated with the emitted quanta, rather than at the divergent
static-observer energy \eqref{eq:static_energy}. The specific identification of $E_\star$ (e.g.\ $E_\infty$,
a regular-frame local energy, or a self-consistent estimate tied to the emission process) is left
deliberately open, but it is held fixed across implementations when comparing results. Under this
assumption, the near-horizon temperature depends only on $g/f$, as shown in
Section~\ref{sec:temperature_scaling}.

%============================================================
\section{Static horizons and the common \texorpdfstring{$g/f$}{g/f} scaling of the Hawking temperature under a shared operational scale}
\label{sec:temperature_scaling}
%============================================================

\subsection{Standard result on an energy-independent background}
For the metric \eqref{eq:static_metric}, the horizon radius $r_h$ is defined by $F(r_h)=0$. The standard
surface gravity is
\begin{equation}
\kappa_0=\frac{1}{2}\,F'(r_h)\,,
\end{equation}
and the corresponding Hawking temperature is \cite{Hawking1975,WaldBook}
\begin{equation}
T_0=\frac{\kappa_0}{2\pi}=\frac{F'(r_h)}{4\pi}.
\label{eq:T0}
\end{equation}
For example, for the Schwarzschild case $F(r)=1-2M/r$, one finds $r_h=2M$ and $T_0=1/(8\pi M)$.

\subsection{Implementation A: local-frame MDR on a fixed background}
We outline the near-horizon Hamilton--Jacobi (HJ) tunneling argument at a level sufficient to isolate the
MDR dependence. Tunneling methods provide a geometrically transparent route to the temperature and are
known to agree with the surface-gravity result for stationary horizons when implemented consistently
\cite{ParikhWilczek2000,Angheben2005,VanzoAcquavivaDiCriscienzo2011}.

Consider a massless mode with HJ action $S=-Et+W(r)+\cdots$, where $E$ denotes the conserved Killing energy
labelling the stationary mode. In an orthonormal frame adapted to the static geometry, one may take
\begin{equation}
p_{\hat t}=\frac{E}{\sqrt{F(r)}},
\qquad
p_{\hat r}=\sqrt{F(r)}\,p_r,
\end{equation}
where hats denote local orthonormal components and $p_r=\partial_r W$. Imposing the massless MDR in the
form \eqref{eq:fg_mdr} yields
\begin{equation}
p_{\hat r}
=
p_{\hat t}\,\frac{f(E_\star/\Epl)}{g(E_\star/\Epl)}
=
\frac{E}{\sqrt{F(r)}}\,\frac{f_\star}{g_\star},
\qquad
f_\star\equiv f(E_\star/\Epl),\ \ g_\star\equiv g(E_\star/\Epl).
\label{eq:pr_hat_relation}
\end{equation}
Here we have implemented the operational prescription adopted throughout: the deformation functions are
evaluated at a finite scale $E_\star$ associated with the emitted quantum, not at the divergent static
energy $E/\sqrt{F(r)}$. In other words, $E_\star$ is regarded as a physical scale determined by the
emission process (or by the detector at infinity) and is treated as finite in the near-horizon computation.

Converting to the coordinate momentum gives
\begin{equation}
p_r=\frac{p_{\hat r}}{\sqrt{F(r)}}=\frac{E}{F(r)}\,\frac{f_\star}{g_\star}.
\label{eq:pr_coordinate}
\end{equation}
Near the horizon, $F(r)\simeq F'(r_h)(r-r_h)$, and the radial integral controlling the imaginary part of
the action develops a simple pole:
\begin{equation}
\mathrm{Im}\,W
=
\mathrm{Im}\int p_r\,\dd r
=
\mathrm{Im}\int \frac{E}{F(r)}\,\frac{f_\star}{g_\star}\,\dd r
=
\frac{\pi E}{F'(r_h)}\,\frac{f_\star}{g_\star},
\end{equation}
where the last equality follows from contour deformation across the pole. Interpreting the tunneling rate
as $\Gamma\propto \exp(-2\,\mathrm{Im}\,S)$ and matching to a Boltzmann factor $\exp(-E/T)$ yields
\begin{equation}
T(E_\star)=\frac{F'(r_h)}{4\pi}\,\frac{g_\star}{f_\star}
= T_0\,\frac{g(E_\star/\Epl)}{f(E_\star/\Epl)}.
\label{eq:T_g_over_f}
\end{equation}
Thus, within this operational prescription, the local-frame MDR modifies the Hawking temperature solely
through the ratio $g/f$.

(i) The derivation highlights that the \emph{pole structure} is geometric (encoded in $F(r)$), whereas the
MDR enters only through the prefactor $f_\star/g_\star$. (ii) If one instead attempted to evaluate
$f$ and $g$ at the divergent $E/\sqrt{F}$, the result would become ill-defined; this reflects the fact
that the operational meaning of the deformation scale must be specified in a way that remains finite for a
regular physical process.

\subsection{Implementation B: rainbow metric and the same scaling}
In rainbow gravity one introduces the energy-dependent orthonormal frames \eqref{eq:rainbow_frames}. For
the static seed metric \eqref{eq:static_metric}, this leads to the effective line element
\begin{equation}
\dd s^2(E_\star)=
-\frac{F(r)}{f_\star^{\,2}}\,\dd t^2
+\frac{1}{g_\star^{\,2}}\frac{\dd r^2}{F(r)}
+\frac{r^2}{g_\star^{\,2}}\,\dd\Omega^2,
\label{eq:rainbow_metric_static}
\end{equation}
with $f_\star=f(E_\star/\Epl)$ and $g_\star=g(E_\star/\Epl)$. The horizon location remains determined by
$F(r_h)=0$, but the relation between redshift and proper distance is modified by $(f_\star,g_\star)$.

A direct computation of the surface gravity illustrates the universal factor. Let $\chi=\partial_t$ be the
stationary Killing vector. Its norm is $\chi^2=g_{tt}(E_\star)=-F/f_\star^2$. For static metrics of the
form \eqref{eq:rainbow_metric_static}, one may use the standard expression
\begin{equation}
\kappa^2(E_\star)=
-\frac{1}{2}\left(\nabla_\mu \chi_\nu\right)\left(\nabla^\mu \chi^\nu\right)\bigg|_{r=r_h},
\end{equation}
or equivalently
\begin{equation}
\kappa(E_\star)=
\frac{1}{2}\left.\frac{\partial_r(-\chi^2)}{\sqrt{-\chi^2\,g_{rr}(E_\star)}}\right|_{r=r_h}.
\end{equation}
Using $-\chi^2=F/f_\star^2$ and $g_{rr}(E_\star)=(g_\star^{-2})F^{-1}$ yields
\begin{equation}
\kappa(E_\star)
=
\frac{1}{2}\left.\frac{F'(r)/f_\star^2}{\sqrt{(F/f_\star^2)\,(1/g_\star^2)(1/F)}}\right|_{r=r_h}
=
\frac{g_\star}{f_\star}\,\frac{1}{2}F'(r_h)
=
\frac{g_\star}{f_\star}\,\kappa_0.
\end{equation}
Therefore the Hawking temperature inferred from Euclidean regularity or from the surface gravity is
\begin{equation}
T(E_\star)=\frac{\kappa(E_\star)}{2\pi}
=
T_0\,\frac{g_\star}{f_\star},
\end{equation}
which coincides with \eqref{eq:T_g_over_f}.

For static, spherically symmetric horizons, the fixed-background local-frame MDR and rainbow-metric
prescriptions coincide at the level of the surface-gravity/tunneling temperature, \emph{provided they are
evaluated at the same finite operational scale} $E_\star$. This equivalence is a central organizing
principle for interpreting and comparing results in DSR/rainbow black-hole thermodynamics.

%============================================================
\section{Representative deformations: AC, MS, and G-DSR/GDRS}
\label{sec:examples}
%============================================================

\subsection{AC-type MDR}
For \eqref{eq:fg_AC_exact}, one has
\begin{equation}
\frac{g_\star}{f_\star}=\sqrt{1-\eta\,\frac{E_\star}{\Epl}},
\end{equation}
and \eqref{eq:T_g_over_f} yields
\begin{equation}
T_{\rm AC}(E_\star)=T_0\sqrt{1-\eta\frac{E_\star}{\Epl}}
\simeq
T_0\left(1-\frac{\eta}{2}\frac{E_\star}{\Epl}\right)
\qquad (E_\star/\Epl\ll 1).
\end{equation}
For $\eta>0$ the temperature is reduced at fixed $E_\star$, whereas for $\eta<0$ it is increased within the
validity domain of the effective expansion. If one adopts an additional prescription that ties $E_\star$ to
the typical emission energy (for instance $E_\star\sim \mathcal{O}(T)$), the correction becomes
mass-dependent in the Schwarzschild case. Such self-consistent prescriptions have been explored in the
rainbow literature in connection with possible remnants \cite{LingHuLi2006,AliFaizalKhalil2014,GalanMenaMarugan2006}.

\subsection{MS invariant}
For \eqref{eq:fg_MS}, $g_\star/f_\star=1$, and therefore
\begin{equation}
T_{\rm MS}(E_\star)=T_0
\end{equation}
at the surface-gravity/tunneling level. This exemplifies a broader point: a non-trivial MDR can leave the
near-horizon temperature unchanged whenever $f=g$ (equivalently, $g/f=1$). Importantly, this does \emph{not}
imply that all aspects of evaporation are unmodified. Even when $T$ is unchanged, the emission spectrum can
be affected by modified densities of states, phase-space measures, and multi-particle kinematics, which are
genuine ingredients of DSR beyond the one-particle mass shell \cite{Hossenfelder2007}.

\subsection{G-DSR/GDRS family}
For \eqref{eq:fg_GDRS} the temperature becomes
\begin{equation}
T_{\mathrm{GDRS}}(E_\star)
=
T_0\sqrt{\frac{1-2\,\Delta\alpha\,(E_\star/\Epl)}{1-2\,\alpha_2\,(E_\star/\Epl)}}
\simeq
T_0\left[1-(\Delta\alpha-\alpha_2)\frac{E_\star}{\Epl}\right],
\label{eq:T_GDRS}
\end{equation}
so that the leading correction is governed by the single combination $\Delta\alpha-\alpha_2$ and vanishes
for the symmetric subfamily $\Delta\alpha=\alpha_2$.

Within the leading-order domain $E_\star/\Epl\ll 1$, the sign of $\Delta\alpha-\alpha_2$ determines whether
the temperature is suppressed or enhanced at fixed $E_\star$. For positive $\Delta\alpha-\alpha_2$ one finds
$T_{\mathrm{GDRS}}<T_0$, while for negative $\Delta\alpha-\alpha_2$ one finds $T_{\mathrm{GDRS}}>T_0$.
Moreover, if one interprets \eqref{eq:fg_GDRS} beyond leading order, reality of $f$ and $g$ would require
$1-2\alpha_2 x>0$ and $1-2\Delta\alpha x>0$; such constraints should be viewed as indicative of the
effective model's domain rather than as fundamental bounds.

\subsection{Illustration for Schwarzschild and simple thermodynamic implications}
To connect the temperature rescaling to thermodynamics, consider a Schwarzschild black hole with
$T_0=1/(8\pi M)$. If one holds $E_\star$ fixed externally (e.g.\ by specifying a detector energy at
infinity), then the temperature shift is simply a multiplicative factor independent of $M$:
\begin{equation}
T(M;E_\star)=\frac{1}{8\pi M}\,\frac{g(E_\star/\Epl)}{f(E_\star/\Epl)}.
\end{equation}
In many physical situations, however, it is natural to relate $E_\star$ to the typical energy of emitted
quanta, which is of order $T$ for a thermal spectrum. A minimal phenomenological choice is
$E_\star=\xi\,T_0$ with $\xi=\mathcal{O}(1)$, which yields (for GDRS at leading order)
\begin{equation}
T(M)\simeq \frac{1}{8\pi M}\left[1-(\Delta\alpha-\alpha_2)\,\xi\,\frac{1}{8\pi M\,\Epl}\right].
\end{equation}
Using the first law $ \dd S = \dd M/T(M)$ then gives, to the same order,
\begin{equation}
S(M)\simeq 4\pi M^2 + 8\pi\,(\Delta\alpha-\alpha_2)\,\frac{\xi}{8\pi\,\Epl}\,M + \text{const.}
\end{equation}
This illustrates two general lessons: (i) once a prescription for $E_\star$ is specified, temperature
corrections translate into entropy corrections, but the functional dependence on $M$ depends on that
prescription; (ii) for macroscopic $M$ the correction is parametrically suppressed by $1/(M\Epl)$, and
hence is extremely small.

%============================================================
\section{Beyond the near-horizon temperature: sources of model dependence}
\label{sec:beyond_temperature}
%============================================================
The common scaling \eqref{eq:T_g_over_f} at the surface-gravity/tunneling level does not imply that all
DSR-motivated models lead to identical evaporation phenomenology. The Hawking temperature captures the
near-horizon analyticity and redshift structure, but the observable flux at infinity depends on additional
inputs. For a generic species $i$ with greybody factor $\Gamma_i(\omega)$, the standard energy flux has the
schematic form
\begin{equation}
\frac{\dd E}{\dd t}\sim \sum_i \int_0^\infty \frac{\dd\omega}{2\pi}\,
\frac{\omega\,\Gamma_i(\omega)}{\exp(\omega/T)\mp 1},
\end{equation}
and each ingredient may receive DSR/rainbow modifications.

Several concrete mechanisms can generate model dependence beyond the common near-horizon scaling.
First, massive thresholds and the relation between energy, momentum, and group velocity can modify the
emission rates even in cases where $T(E_\star)$ itself is unchanged, such as models with $f=g$.
Second, DSR constructions with curved momentum space can induce a non-trivial integration measure and
therefore alter the density of states and the spectral distribution
\cite{Hossenfelder2007,BarcaroliPfeifer2015}. Third, the greybody factors $\Gamma_i(\omega)$ can change if
the effective wave equation, turning points, or propagation speed in the exterior region are modified;
this effect can easily compete with or dominate a small temperature rescaling. Finally, in full DSR one
expects non-linear composition laws for multi-particle kinematics, so the bookkeeping of emitted energy
and the implementation of backreaction need not coincide with the standard semiclassical picture.
These observations explain why agreement at the level of the near-horizon temperature does not imply
agreement for integrated evaporation histories or for endpoint behavior.

The operational choice of $E_\star$ remains the dominant ambiguity even inside the temperature calculation
itself. If $E_\star$ is identified with a detector energy at infinity, the deformation acts mainly as a
frequency-dependent reweighting of the observed spectrum. If instead one ties $E_\star$ self-consistently
to the local emission process, for example by taking $E_\star\sim T$, the correction becomes mass dependent
and can qualitatively alter late-stage evaporation in models where $g/f$ approaches zero at finite
argument \cite{GalanMenaMarugan2006,AliFaizalKhalil2014}. For the common estimate $E_\star=\xi T_0$ in the
GDRS family, the relative correction is
\begin{equation}
\frac{\delta T}{T_0}\simeq -(\Delta\alpha-\alpha_2)\,\frac{\xi}{8\pi M\Epl}.
\end{equation}
Thus the effect is suppressed by the inverse black-hole mass in Planck units. As an order-of-magnitude
illustration, one finds $|\delta T|/T_0\sim 4\times 10^{-40}|\Delta\alpha-\alpha_2|\,\xi$ for a solar-mass
black hole, whereas even a much lighter black hole of mass $10^{15}\,$g still gives only
$|\delta T|/T_0\sim 10^{-21}|\Delta\alpha-\alpha_2|\,\xi$. Any phenomenological bound derived from Hawking
thermodynamics would therefore be weak unless one has access to primordial or genuinely Planckian black
holes. This is precisely why a meaningful comparison across approaches must specify not only the
prescription for $E_\star$ but also the dynamical and statistical ingredients beyond the near-horizon
temperature.

%============================================================
\section{Discussion: implications of the G-DSR/GDRS extension and interpretation of results}
\label{sec:gdrs_discussion}
%============================================================
The generalized DSR family \eqref{eq:gdrs_mdr_raw}--\eqref{eq:fg_GDRS} provides a compact two-parameter
organization of leading-order Planck-suppressed deformations and clarifies which parameter combinations
control the Hawking-temperature rescaling. Within the present assumptions (static horizons and a fixed
finite $E_\star$), its imprint on the temperature is governed solely by $g/f$, and hence by the difference
$\Delta\alpha-\alpha_2$ at leading order; see \eqref{eq:T_GDRS}. Several points merit emphasis.

Different microscopic realizations that share the same value of $\Delta\alpha-\alpha_2$ are
thermodynamically degenerate at $\mathcal{O}(E_\star/\Epl)$ with respect to the near-horizon temperature.
Thus, at this order, the mapping from ``microphysics'' to temperature corrections is not injective: one
expects additional observables (e.g.\ dispersion-induced group velocities, measures, or composition laws)
to be required to break degeneracies.

The symmetric subfamily $\Delta\alpha=\alpha_2$ yields $g/f=1$ and therefore $T(E_\star)=T_0$ at leading
order. This generalizes the MS observation and highlights a structural point: if the deformation preserves
the equality between temporal and spatial rainbow functions, the temperature is unchanged at the
surface-gravity level. Such models may be of special interest in light of locality considerations for
energy-dependent propagation in DSR-like frameworks \cite{Hossenfelder2010}: deformations that do not
induce an energy-dependent speed for massless quanta (often correlated with $f=g$) can evade the strongest
constraints arising from non-locality arguments, although a complete assessment depends on the full
multi-particle formulation.

For macroscopic black holes, any reasonable prescription yields $E_\star/\Epl\ll 1$. If $E_\star$ is taken
to be of order the Hawking temperature itself, then $E_\star/\Epl\sim T_0/\Epl\sim (M\Epl)^{-1}\ll 1$ for
$M\gg \mpl$. Consequently, leading-order GDRS corrections are extremely small for astrophysical black holes
and become potentially relevant only as $M$ approaches the Planck mass. In that regime, however, the
semiclassical approximation underlying both the tunneling method and rainbow effective metrics becomes
suspect, and one expects backreaction and genuinely quantum-gravitational dynamics to dominate the
endpoint.

Our main technical result may be interpreted as follows. For static horizons, the temperature is governed
by two ingredients: (i) the geometric pole structure in the radial momentum near the horizon, and (ii) the
relation between energy and radial momentum implied by the MDR. Under the operational assumption that the
deformation functions are evaluated at a fixed finite scale $E_\star$, ingredient (ii) enters only as an
overall rescaling. Rainbow metrics implement precisely the same rescaling through an energy-dependent
normalization of time and space in the orthonormal frame. Therefore, at the level of near-horizon
temperature and under the shared-scale assumption, the two prescriptions are equivalent parametrizations.

The analysis is intentionally restricted to static, spherically symmetric horizons. Extensions to rotating
or dynamical horizons require additional care: the definition of energy scales, the role of non-trivial
horizon generators, and the use of Kodama-like vectors in non-stationary settings can all influence the
operational meaning of $E_\star$. Moreover, the full evaporation problem requires a specification of the
spectral distribution, greybody factors, and the consistent implementation of energy conservation and
composition laws. These extensions are natural directions for future work within the G-DSR/GDRS
framework.

%============================================================
\section{Conclusion}
%============================================================
We have compared two standard prescriptions for implementing DSR-motivated modified dispersion relations
in black-hole spacetimes: (A) an energy-independent background with the MDR imposed in local orthonormal
frames, and (B) the rainbow-metric construction in which the deformation is encoded in an explicitly
energy-dependent effective geometry. The main objective of the paper has been to determine, in a precise
and controlled way, whether these two prescriptions imply genuinely distinct near-horizon Hawking-temperature
predictions or whether their apparent difference is mainly a matter of formal representation. For static,
spherically symmetric horizons, we showed explicitly that both prescriptions lead to the same near-horizon
modification of the Hawking temperature at the surface-gravity/tunneling level when expressed in the
$f$--$g$ parametrization and evaluated at the same finite operational scale $E_\star$,
\begin{equation}
T(E_\star)=T_0\,\frac{g(E_\star/\Epl)}{f(E_\star/\Epl)}\,,\qquad T_0=\frac{\kappa_0}{2\pi}\,.
\end{equation}
Therefore, the main conclusion should be read as a conditional equivalence statement: once the physical
prescription for $E_\star$ is fixed and shared, the distinction between ``local-frame MDR on a fixed
background'' and ``rainbow metric'' becomes largely one of parametrization at the level of the
near-horizon temperature.

Applying the general scaling to representative deformations, we found that an Amelino--Camelia-type MDR
suppresses the temperature at fixed $E_\star$ for $\eta>0$, whereas the original Magueijo--Smolin invariant
satisfies $f=g$ and leaves the surface-gravity temperature unchanged. We also incorporated the generalized
DSR extension (G-DSR/GDRS) characterized at leading order by $(\alpha_2,\Delta\alpha)$
\cite{BoumaliJafariChargui2026}, for which the correction is controlled by the single combination
$\Delta\alpha-\alpha_2$ and vanishes on the symmetric subfamily $\Delta\alpha=\alpha_2$.

Just as importantly, the analysis makes clear what has \emph{not} been established. The operational
identification of the deformation scale $E_\star$ is not derived from first principles here; rather, it is
an input that must be specified before the comparison can be carried out. Likewise, agreement in the
near-horizon temperature does not fix the evaporation phenomenology, because thresholds, phase-space
measures, greybody factors, and non-linear composition laws can all introduce additional model dependence.
For the common choice $E_\star\sim T_0$, the relative correction is suppressed by $1/(M\Epl)$ and is
therefore tiny for macroscopic black holes, so any realistic phenomenological impact is expected only in
primordial or nearly Planckian regimes where semiclassical control is itself limited.

Taken together, these results show that the present work should be viewed as a clarification of how two
common curved-spacetime DSR implementations are related, together with a compact parametrization of the
leading G-DSR/GDRS correction. Nevertheless, a fully predictive phenomenology will require going beyond the
near-horizon temperature and deriving or motivating the relevant operational energy prescription in a more
fundamental way.

%============================================================

\end{document}